\documentclass[usegraphicx,usenatbib]{mn2e}
\usepackage{color}

\newcommand{\simgt}{\lower.5ex\hbox{$\; \buildrel > \over \sim \;$}}
\newcommand{\simlt}{\lower.5ex\hbox{$\; \buildrel < \over \sim \;$}}

\begin{document}
\title[Shear Power Spectrum Reconstruction]{Shear Power Spectrum Reconstruction using Pseudo-Spectrum Method}

\author[Hikage et al.]{Chiaki Hikage$^1$, Masahiro Takada$^2$, Takashi Hamana$^3$, David Spergel$^{1,2}$ \\
$^1$ Department of Astrophysical Sciences, Princeton University, Peyton Hall, Princeton NJ 08544, USA \\
$^2$ Institute for the Physics and Mathematics of the Universe (IPMU), 
The University of Tokyo, Chiba 277-8582, Japan \\
$^3$ National Astronomical Observatory of Japan, Tokyo 181-8588, Japan}
\maketitle

\begin{abstract}
We develop a pseudo power spectrum technique for measuring the lensing
power spectrum from weak lensing surveys in both the full sky and flat
sky limits. The power spectrum approaches have a number of advantages
over the traditional correlation function approach.  We test the
pseudo spectrum method by using numerical simulations with
square-shape boundary that include masked regions with complex
configuration due to bright stars and saturated spikes.  Even when
25\% of total area of the survey is masked, the method recovers the
$E$-mode power spectrum at a sub-percent precision over a wide range
of multipoles $100\simlt\ell\simlt 10^4$. The systematic error is
smaller than the statistical errors expected for a 2000 square degree
survey. The residual $B$-mode spectrum is well suppressed in the
amplitudes at less than a percent level relative to the $E$-mode.  We
also find that the correlated errors of binned power spectra caused by
the survey geometry effects are not significant.  Our method is
applicable to the current and upcoming wide-field lensing surveys.
\end{abstract}

\section{Introduction}
\label{sec:intro}

Large scale structure deflects light rays as they propagate from
distant galaxies to us, thus distorting the shapes of these galaxies
\citep[e.g.,][for thorough reviews]{BS01,HoekstraJain08}. This weak
lensing or cosmic shear signals measures a combination of the total
matter distribution projected along the line-of-sight and the angular
diameter distance. Since the first measurements of weak lensing only a
decade ago \citep{VW00,Wittman00,Bacon00,Kaiser00}, there have been
major improvements as surveys continue to grow in size and depth
\citep[e.g.][for the latest
results]{Fuetal08,Ichiki09,Schrabback10}. Gravitational lensing is one
of the most promising methods of constraining cosmology including the
nature of dark energy \citep[e.g.][]{TakadaBridle07}. There are various
on-going and planned surveys aimed at studying dark energy through the
high-precision weak lensing measurements: the CFHT Legacy
Survey\footnote{http://www.cfht.hawaii.edu/Science/CFHLS/}, the Hyper
Suprime-Cam Weak Lensing Survey \citep{Miyazaki06}, the Dark Energy
Survey (DES)\footnote{http://www.darkenergysurvey.org}, and ultimately
Large Synoptic Survey Telescope (LSST)\footnote{http://www.lsst.org},
Euclid \citep{Refregier10}, and Joint Dark Energy Mission (JDEM).

How should we analyze these new lensing data set?  Most researchers
use the two-point correlation function to characterize the cosmic
shear signals. The correlation function method can be easily applied
to complex survey geometries involving partial sky coverage and masked
regions. However, the errors in the measurement are highly correlated
between different bins \citep[see][for the detailed
studies]{Schneider02,Joachimietal08}. Even if the shear field follows
Gaussian statistics, which is a good approximation in the linear
regime, there are large correlations between different angular-scale
bins.  These correlations are even larger, and more model-dependent on
the smaller scales that contain most of the current observational
information. Since the accurate estimate of the covariances is
essential for robust cosmological constraints, a large number of
numerical simulations are necessary \citep{Semboloni07}.

The power spectrum, the Fourier- or Harmonic-transformed counterpart
of the two-point correlation function, is an alternative means of
measuring the cosmic shear correlations.  Whilst the correlation
function and power spectrum are mathematically equivalent, the power
spectrum measurement of cosmic shear has been used less \citep[see the
COMBO 17-survey by][]{Brown03}. The power spectrum approach has a
number of advantages: its theoretical interpretation is simpler and
there are weaker correlations between band powers at different
multipoles. For example, the different bins are independent for the
Gaussian field or on large angular scales. Even for small angular
scales affected by nonlinear structure formation, the power spectrum
covariances are relatively well understood through both analytical
models and simulations of nonlinear structure formation
\citep{HuWhite01,CoorayHu01,TakadaJain09,Sato09,Pielorzetal09}.  The
disadvantage is the presence of finite sky coverage and masked regions,
which breaks the orthogonality of Fourier/Harmonic components.  One
needs to properly deal with the survey geometry effect to estimate
unbiased power spectrum.

The purpose of this paper is to eliminate this disadvantage.  We
employ the {\em pseudo} power spectrum technique, which is well
developed in the CMB studies \citep[e.g.][]{Wandelt01} \footnote{See
\citet{Seljak98} and \citet{HuWhite01} for the maximum likelihood
method of shear power spectrum estimation.}. For the first time, we
apply the method to recover the lensing power spectrum from the shear
field taking into account incomplete survey geometry.  To assess the
performance of this method, we make simulated shear maps including a
realistic configuration of masked regions due to bright stars and
saturated spikes. Furthermore, we develop the method for both the
full-sky and flat-sky approaches.  The full-sky approach is adequate
for reconstructing large angular-scale modes that are relevant for the
curvature of the sky.  On the other hand, the flat-sky approach should
serve as a practically useful approximation of sub-degree scale modes,
which carry most of useful cosmological information in the shear power
spectrum.  We find that the pseudo power spectrum method allows for an
unbiased estimate of the underlying $E$-mode power spectrum over a
range of angular scales we study. We also show that the residual
$B$-mode power spectrum, which is leaked from $E$-mode power due to an
imperfect reconstruction, can be well suppressed. Our method can be
applied to the existing data and forthcoming weak lensing surveys.

The paper is organized as follows: Section \ref{sec:method} describes
the pseudo spectrum method to deconvolve shear power spectra with
inhomogeneous survey mask.  Section \ref{sec:simulations} describes
the simulation maps we use to test the deconvolution method. We employ
two different simulation maps: one is Gaussian shear maps and the
other is the ray-tracing simulations of shear maps including the
non-Gaussian effects due to nonlinear structure formation.  Section
\ref{sec:results} shows the results of both the full-sky and flat-sky
approaches.  Section~\ref{sec:summary} is devoted to the summary and
conclusions.

\section{Methodology: Reconstruction of Shear Power Spectrum}
\label{sec:method}

In this section we briefly review a method for reconstructing shear
power spectra from the pseudo-spectrum estimators. We take into
account an imperfect survey geometry due to survey boundary and
masking effect.  The method is analogous to the one used in estimating
CMB polarization power spectra \citep{Kogut03,Brown05}.

\subsection{Full Sky Formalism}
\label{subsec:full}

Since the shear field is a spin-2 field, the $E$- and $B$-mode
harmonic coefficients of the shear fields $\gamma_i (i=1,2)$ can be
expressed in the spherical harmonic expansion as
\begin{equation}
\label{eq:def_eb}
E_{lm}\pm iB_{lm}=\oint\! d\Omega_{\hat\mathbf{n}} \left[
\gamma_1(\hat{\mathbf n})\pm i\gamma_2(\hat{\mathbf n})
\right]
~ {}_{\pm 2}Y_{lm}^{\ast}(\hat{\mathbf n}),
\end{equation}
and the inverse relation is
\begin{equation}
\label{eq:def_eb2}
\gamma_1(\hat{\mathbf n})\pm i\gamma_2(\hat{\mathbf n}) =
\sum_{lm} \left[ E_{lm}\pm iB_{lm} \right] ~ {}_{\pm 2}Y_{lm}(\hat{\mathbf n}),
\end{equation}
where ${}_{\pm 2}Y_{lm}$ is the spin-2 spherical harmonics, and
$\hat{\mathbf n}$ denotes the unit vector specifying the angular
direction on the sky. The integration range is over the full sky.

In the linear regime, the statistical information in the map is fully
encoded in the power spectra
\begin{eqnarray}
C_l^{EE}&\equiv &\frac{1}{2l+1}\sum_m|E_{lm}|^2, 
\label{eq:ce_def}\\
C_l^{BB}&\equiv &\frac{1}{2l+1}\sum_m|B_{lm}|^2, 
\label{eq:cb_def}\\
C_l^{EB}&\equiv &\frac{1}{2l+1}\sum_m E_{lm}B_{lm}^\ast.
\label{eq:ceb_def}
\end{eqnarray}
In the single lens limit, the shear field arising from a scalar
gravitational field should be a gradient or curl-free field
($B_{lm}=0$). The multiple lensing effect generates B-mode power
spectrum, but their power is $\sim 10^4$ times smaller than the E-mode
power. Thus the $E$-mode power spectrum effectively contains all
information on the cosmic shear, i.e. $E$-mode power spectrum is
equivalent to that of the projected mass density along the
line-of-sight between source galaxies and an observer. Hence the
$B$-mode can be used as a monitor of residual systematic effects.  The
standard methods to separate $E/B$ mode correlation functions involve
integrals of the measured correlation functions down to arbitrary
small scale or up to very large scale. As the scale range accessible
from a finite sky data is limited, residual uncertainties are generated
\citep[][also see \citealt{Schneideretal10} for a new method using the
limited-range integration of correlation function to separate the
$E$-mode]{Schneideretal98,Crittenden02}.

Observational effects, such as a finite sky coverage and bright star
masks, limit the survey area to a region $K(\hat{\mathbf n})$.  The
observed shear field is modified as
\begin{equation}
\tilde{\gamma}_1(\hat{\mathbf n})\pm i\tilde{\gamma}_2(\hat{\mathbf n})
=K(\hat{\mathbf n})(\gamma_1(\hat{\mathbf n})\pm i\gamma_2(\hat{\mathbf n})).
\label{eq:obsshear}
\end{equation}
Without weighting, $K(\hat{\mathbf{n}})=0$ if the position vector
$\hat{ \mathbf{n}}$ lies in masked regions or regions outside the
survey, otherwise $K(\hat{\mathbf{n}})=1$ within the survey.  This
finite sky coverage couples modes and generate artificial $B$ modes.  We
can describe the observed shear fields in terms of ``pseudo E and B
modes'', denoted as $\tilde{E}_{lm}$ and $\tilde{B}_{lm}$
\begin{equation}
\tilde{E}_{lm}\pm i\tilde{B}_{lm}=\oint d\Omega_{\hat\mathbf{n}} 
\left[
K(\hat{\mathbf n})(\gamma_1(\hat{\mathbf n})\pm i\gamma_2(\hat{\mathbf n}))
\right]
{}_{\pm 2}Y_{lm}^{\ast}(\hat{\mathbf n}).
\end{equation}
These pseudo E and B modes are related to the true E and B modes as
\begin{equation}
\tilde{E}_{lm}\pm i\tilde{B}_{lm}=\sum_{l' m'}
(E_{l' m'} \pm iB_{l' m'}){}_{\pm 2}
W_{ll' mm'},
\label{eq:ebcoeff_mask}
\end{equation}
through a convolution kernel,
\begin{eqnarray}
{}_{\pm 2}W_{ll' mm'}&\equiv &
\oint d\Omega_{\hat{\mathbf n}}~ 
{}_{\pm 2}Y_{l' m'}(\hat{\mathbf n})
K(\hat{\mathbf n}){}_{\pm 2}Y_{lm}^\ast(\hat{\mathbf n}) \nonumber \\
&&\hspace{-3em}=
\sum_{l''m''}K_{l''m''}
(-1)^m\sqrt{\frac{(2l+1)(2l'+1)(2l''+1)}{4\pi}}
\nonumber \\ &\times &
\left(\begin{array}{ccc}
l & l' & l'' \\
\pm 2 & \mp 2 & 0
\end{array}\right)
\left(\begin{array}{ccc}
l & l' & l'' \\
m & m' & m''
\end{array}\right),
\end{eqnarray}
where $\left(\begin{array}{ccc}
l_1 & l_2 & l_3 \\
m_1 & m_2 & m_3
\end{array}\right)$ are known as the Wigner $3j$ symbols
\citep[see the references in][]{DahlenTromp} and $K_{lm}$ is the
harmonic transform of the mask function $K(\hat\mathbf{n})$:
\begin{equation}
K_{lm}=\oint d{\Omega}_{\hat{\mathbf n}} 
K(\hat{\mathbf n})Y_{lm}^\ast(\hat{\mathbf
n}).
\end{equation}

The pseudo power spectra $\tilde{C}_l^{EE}, \tilde{C}_l^{BB} $ and $
\tilde{C}_l^{EB}$ are defined similarly to the
equations~(\ref{eq:ce_def})-(\ref{eq:ceb_def}) using pseudo $E$ and
$B$ modes (eq.[\ref{eq:ebcoeff_mask}]). After straightforward
algebraic calculation, one can find that the pseudo and true spectra
are related to each other via
\begin{equation}
\label{eq:pseudo}
\tilde{\mathbf C}_l =\sum_{l'}{\mathbf
M}_{ll'}F_{l'}^2{\mathbf C}_{l'} +\tilde{\mathbf
N}_l,
\end{equation}
where we have introduced the vector notations
$\tilde{\mathbf{C}}_l\equiv (\tilde{C}_l^{EE}, \tilde{C}_l^{BB},
\tilde{C}_l^{EB})$ and so on for notational simplicity, and $F_l$ is
the pixel window function. In the above equation we include the shot
noise contribution arising from the intrinsic ellipticities of source
galaxies. The intrinsic noise is simply modeled as the convolved noise
power spectrum $\tilde\mathbf{N}_l$ with the pixel window and the
mask.  Non-zero components of the mode-mode coupling matrix
$\mathbf{M}_{ll'}$ are given as
\begin{eqnarray}
M_{ll'}^{EE,EE}
&=&M_{ll'}^{BB,BB} \nonumber \\
&=& \frac{2l' +1}{8\pi}\sum_{l''}
(2l''+1){\cal K}_{l''}
[1+(-1)^{l+l'+l''}] \nonumber \\
& \times &
\left(\begin{array}{ccc}
l & l' & l''\\
2 & -2 & 0 \\
\end{array}\right)^2, \\
M_{ll'}^{EE,BB}
&=&M_{ll'}^{BB,EE} \nonumber \\
&=& \frac{2l' +1}{8\pi}\sum_{l''}(2l''+1){\cal K}_{l''}
[1-(-1)^{l+l'+l''}] \nonumber \\
&\times &
\left(\begin{array}{ccc}
l & l' & l''\\
2 & -2 & 0 \\
\end{array}\right)^2, \\
M_{ll'}^{EB,EB}
&=& \frac{2l' +1}{4\pi}\sum_{l''}
(2l''+1){\cal K}_{l''}
\left(\begin{array}{ccc}
l & l' & l''\\
2 & -2 & 0 \\
\end{array}\right)^2 , \nonumber \\
\end{eqnarray}
with ${\cal K}_l$ being defined as
\begin{equation}
{\cal K}_l\equiv\frac{1}{2l+1}\sum_m K_{lm} K_{lm}^\ast .
\end{equation}
Note that $M_{ll'}^{EE,EB}=0=M_{ll'}^{BB,EB}$.  Equation
(\ref{eq:pseudo}) tells that an imperfect survey geometry causes
mode-mixing or equivalently a leakage of $E$-mode power into the
$B$-mode even if $C_l^{BB}=0$. 

The underlying power spectra can be reconstructed by solving the
equation (\ref{eq:pseudo}) inversely. The resolution in multipole
space is limited by the survey area, i.e. the finer binning less than
$l_{\rm f}\equiv\sqrt{\pi/f_{\rm sky}}$ does not improve the
statistical significance of the power spectrum reconstruction.  Since
the lensing power spectrum does not have fine scale structures in
multipole space, such a coarse binning is sufficient to capture the
shape of shear power spectrum. Also the coarse binning significantly
reduces the computational cost of measuring the power spectrum in a
wide range of multipoles. We therefore measure the binned power
spectra defined as
\begin{equation}
\mathbf{\cal C}_b\equiv\sum_l^{l\in b} P_{bl}{\mathbf C}_l,
\end{equation}
where the index ``$b$'' denotes the $b$-th multipole bin, and
$\sum_l^{l\in b}$ represents the summation over $l$ between $l_{\rm
min}^{(b)}$ and $l_{\rm min}^{(b+1)}-1$. Here we use a binned operator
$P_{bl}$ so that the binned power becomes the average of dimensionless
power over $l$ between $l_{\rm min}^{(b)}$ to $l_{\rm min}^{(b+1)}-1$
\begin{equation}
P_{bl}\equiv\frac{l(l+1)}{2\pi}\frac{1}{l_{\rm min}^{(b+1)}-l_{\rm min}^{(b)}}.
\end{equation}
The deconvolved binned spectrum is obtained as
\begin{equation}
\mathbf{\cal C}_b = (\mathbf{M}^{-1})_{bb'} \sum_l^{l\in
b'}P_{b' l}(\tilde{\mathbf C}_l-
\langle\tilde{\mathbf N}_l\rangle_{\rm MC}).
\label{eq:deconvolvedCl}
\end{equation}
The mode mixing matrix for the binned spectra is
\begin{equation}
\mathbf{M}_{bb'}=\sum_l^{l\in b}P_{bl}
\sum_{l'}^{l'\in b'}
\mathbf{M}_{ll'}F^2_{l'}Q_{l' b'},
\end{equation}
where $Q_{lb}$ is the reciprocal of $P_{bl}$
\begin{equation}
Q_{lb}\equiv\frac{2\pi}{l(l+1)}.
\end{equation}
The equation~(\ref{eq:deconvolvedCl}) is the key equation for reconstructing
the shear power spectra from masked shear maps in the full-sky
approach in Section \ref{sec:results}.

\subsection{Flat-Sky Approximation}
\label{subsec:flat} 

In this subsection, we present a formalism of a pseudo-spectrum method
in the flat-sky approximation.  The flat-sky approximation is enough
applicable to current lensing survey, such as the CFHT survey covering
a sky of about 200 square degrees in its 4 survey regions
\citep[e.g.][]{Fuetal08}.

In the flat-sky approximation, $E$ and $B$ modes are defined as
\begin{equation}
E_{\mathbf k}^{\rm flat}\pm iB_{\mathbf k}^{\rm flat}
=\int\!d\Omega_{\mathbf n} \left[
\gamma_1({\mathbf n}) \pm i\gamma_2({\mathbf n}) \right]
e^{-i({\mathbf k}\cdot{{\mathbf n}}\pm 2\varphi_{\mathbf k})},
\end{equation}
and inversely related as
\begin{equation}
\gamma_1({\mathbf n}) \pm i\gamma_2({\mathbf n}) 
=\int\frac{d^2 {\mathbf k}}{(2\pi)^2} \left[
E_{\mathbf k}^{\rm flat}\pm iB_{\mathbf k}^{\rm flat}
\right]
e^{i({\mathbf k}\cdot{{\mathbf n}}\pm 2\varphi_{\mathbf k})},
\end{equation}
where the vector ${\mathbf n}$ is a flat-space two-dimensional vector
that approximates the three-dimensional vector $\hat{\mathbf n}$ in
Eq.~(\ref{eq:def_eb}) around some reference point; e.g., if the
coordinate origin in flat space is taken as the north pole, the
position vector of an arbitrary point in the vicinity of the north
pole is specified by two-dimensional vector ${\mathbf
n}=\theta(\cos\varphi,\sin\varphi)=(\theta_x,\theta_y)$. The vector
${\mathbf k}$ is the corresponding wavenumber in the flat-space
coordinate, and $\varphi_{\mathbf k}$ is defined as ${\mathbf
k}=k(\cos\varphi_{\mathbf k}, \sin\varphi_{\mathbf k})$.

Like in the full-sky formalism, we again define pseudo E and B modes
in a flat-sky limit as
\begin{equation}
\tilde{E}_{\mathbf k}^{\rm flat}\pm i\tilde{B}_{\mathbf k}^{\rm flat}
=\int d\Omega_\mathbf{n}
\left[
W(\mathbf{n})(\gamma_1(\mathbf{n})\pm i\gamma_2(\mathbf{n}))
\right]
e^{i({\mathbf k}\cdot{{\mathbf n}}\pm 2\varphi_{\mathbf k})},
\end{equation}
where $W$ represents an arbitrary mask field. We take into account the
mask effect in two steps: a square boundary covering observed survey
regions and the mask inside the square. Even for a complex survey
geometry, such a square shape geometry can be obtained by using the
zero-padding method such that the square-shape region encloses the
whole region of data.

The pseudo $E/B$-mode coefficient $\hat{E}_\mathbf{k}^{\rm flat}\pm
i\hat{B}_\mathbf{k}^{\rm flat}$ for a square field of a side length
$L$ relate to the underlying true coefficients $E_\mathbf{k}^{\rm
flat}\pm i B_\mathbf{k}^{\rm flat}$ as \citep{Bunn02}
\begin{equation}
\hat{E}_{\mathbf{k}}^{\rm flat}\pm i\hat{B}_{\mathbf{k}}^{\rm flat}
=\int\!\!\frac{d^2\mathbf{k'}}{(2\pi)^2}
(E_{\mathbf k'}^{\rm flat}\pm i B_{\mathbf k'}^{\rm flat}) 
S_{\mathbf{k}-\mathbf{k'}}
e^{\pm 2i\varphi_\mathbf{k'}},
\label{eq:ebcoeff_flat}
\end{equation}
where $S_\mathbf{k}$ is the Fourier transform of the survey window
function defined as
\begin{eqnarray}
S_\mathbf{k}
=\frac{\sin (k_xL/2)}{k_x/2}\frac{\sin (k_yL/2)}{k_y/2},
\end{eqnarray}
and $k_x(k_y)$ are $x(y)$-components of $\mathbf{k}$. The window
function $S_{\mathbf k}$ approaches $L^2$ as $\mathbf{k}$ goes to
$\mathbf{0}$.

The shear power spectra in flat-sky approximation is defined as
\begin{eqnarray}
\langle E_{\mathbf k}^{\rm flat}E_{\mathbf k'}^{\rm flat\ast} \rangle &\equiv& 
 (2\pi)^2\delta_D^2(\mathbf{k+k'})C_k^{EE} \nonumber \\
\langle B_{\mathbf k}^{\rm flat}B_{\mathbf k'}^{\rm flat\ast} \rangle &\equiv& 
 (2\pi)^2\delta_D^2(\mathbf{k+k'})C_k^{BB} \nonumber \\
\langle E_{\mathbf k}^{\rm flat}B_{\mathbf k'}^{\rm flat\ast} \rangle &\equiv& 
 (2\pi)^2\delta_D^2(\mathbf{k+k'})C_k^{EB}
\end{eqnarray}
Here $\delta_D^2({\mathbf k})$ is the two-dimensional Dirac delta
function.  From Eq.~(\ref{eq:ebcoeff_flat}) the pseudo power spectra
for the square area relate to the true one as
\begin{eqnarray}
\hat{C}_{\mathbf k}^{EE}&\equiv&
L^{-2}\langle |\hat{E}_{\mathbf k}^{\rm flat}|^2\rangle \nonumber \\
&=&\int\!\!\frac{d^2\mathbf{k'}}{(2\pi)^2}
{\cal S}_{\mathbf{k}\mathbf{-k'}}[
\cos^2(2\varphi_\mathbf{k' k})C_{k'}^{EE}+
\sin^2(2\varphi_\mathbf{k' k})C_{k'}^{BB}],
\nonumber \\
\hat{C}_{\mathbf k}^{BB}&\equiv&
L^{-2}\langle |\hat{B}_{\mathbf k}^{\rm flat}|^2\rangle \nonumber \\
&=&\int\!\!\frac{d^2\mathbf{k'}}{(2\pi)^2}
{\cal S}_{\mathbf{k}\mathbf{-k'}}[
\sin^2(2\varphi_\mathbf{k' k})C_{k'}^{EE}+
\cos^2(2\varphi_\mathbf{k' k})C_{k'}^{BB}],
\nonumber \\
\hat{C}_{\mathbf k}^{EB}&\equiv&
L^{-2}\langle \hat{E}_{\mathbf k}^{\rm flat}\hat{B}_\mathbf{-k}^{\rm flat}\rangle
\nonumber \\
&=&\int\!\!\frac{d^2\mathbf{k'}}{(2\pi)^2}
{\cal S}_{\mathbf{k}\mathbf{-k'}}[
\cos^2(2\varphi_\mathbf{k' k})-\sin^2(2\varphi_\mathbf{k' k})]
C_{k'}^{EB}, \nonumber \\
\label{eq:flat_pseudo}
\end{eqnarray}
where $\varphi_\mathbf{k' k}\equiv
\varphi_\mathbf{k'}-\varphi_\mathbf{k}$ and $\cal{S}_{\mathbf
k}$ is the power spectrum of $S_\mathbf{k}$ defined as
\begin{equation}
\label{eq:pow_square}
\langle S_\mathbf{k}S_\mathbf{k}^\ast\rangle\equiv
 (2\pi)^2\delta_D^2(\mathbf{k+k'}){\cal S}_{\mathbf k}.
\end{equation}
The pseudo spectrum in the equation~(\ref{eq:flat_pseudo}) depend on
wavevector $\mathbf{k}$ rather than its length. We later introduce the
azimuthal angle average over $\varphi_{\mathbf k}$ in reconstructing
the true spectra $C_k$ that depend only on the length of wavevector
$\mathbf{k}$.

Although the relations between the pseudo and underlying spectra
involve an infinite-range integral, we find that the
equation~(\ref{eq:flat_pseudo}) can be approximated by the following
algebraical relations via the mode-mixing matrix:
\begin{equation}
\label{eq:flat_pseudocl1}
\hat{\mathbf C}_{\mathbf k}^{\rm flat} =\sum_\mathbf{k^\prime}{\mathbf
M}^{\cal S}_{\mathbf{kk'}} {\mathbf C}_{k'}^{\rm flat}.
\label{eq:flat_modeS}
\end{equation}
Non-zero components of the mode-mixing matrix are given using the
power spectrum of the power spectrum of survey window
$\cal{S}_{\mathbf k}$ (eq.[\ref{eq:pow_square}]) as
\begin{eqnarray}
M^{{\cal S} 
\it{EE,EE}}_\mathbf{kk'}&=& M_\mathbf{kk'}^{
{\cal S} \it{BB,BB}}\simeq
\overline{\cal S}_{\mathbf{kk'}}^{\rm (cos2)}, \nonumber\\
M^{{\cal S}\it{EE,BB}}_\mathbf{kk'}&=&
 M_\mathbf{kk'}^{{\cal S}\it{BB,EE}}\simeq
\overline{\cal S}_{\mathbf{kk'}}^{\rm (sin2)}, \nonumber\\
M^{{\cal S}\it{EB,EB}}_\mathbf{kk'}&\simeq&
\overline{\cal S}_{\mathbf{kk'}}^{\rm (cos2)} 
-\overline{\cal S}_{\mathbf{kk'}}^{\rm (sin2)},
\end{eqnarray}
where
\begin{eqnarray}
\overline{\cal S}_\mathbf{kk'}^{\rm (cos2)}&=&
\int_{k'_x-\pi/L}^{k'_{x}+\pi/L}\!\frac{dk''_x}{2\pi}
\int_{k'_y-\pi/L}^{k'_y+\pi/L}\!\frac{dk''_y}{2\pi}
\cal{S}_\mathbf{k-k''}\cos^{\rm 2}(\rm{2\varphi}_\mathbf{k''k}),
\nonumber \\ 
\overline{\cal S}_\mathbf{kk'}^{\rm (sin2)}&=&
\int_{k'_x-\pi/L}^{k'_x+\pi/L}\!\frac{dk''_x}{2\pi}
\int_{k'_y-\pi/L}^{k'_y+\pi/L}\!\frac{dk''_y}{2\pi}
\cal{S}_\mathbf{k-k''}\sin^{\rm 2}(\rm{2\varphi}_\mathbf{k''k}).
\nonumber \\
\end{eqnarray}
Note $M^{\cal{S} \it{EE,EB}}=0=M^{\cal{S} \it{BB,EB}}$ as in the
full-sky approach. Since the discrete Fourier decomposition is limited
by the resolution of survey size $L$, we approximate the mode-mixing
matrix as an average of the survey window function multiplied with
either of $\cos^2(2\varphi_\mathbf{kk'})$ or
$\sin^2(2\varphi_\mathbf{kk'})$ over a single pixel of area
$(2\pi/L)^2$ around the vector $\mathbf{k^\prime}$ in Fourier
space. The mode-mixing due to the survey window is important
particularly for large-scale modes ($k\sim 2\pi/L$).

We further need to include the masking effect inside the square
boundary. As in the full-sky approach, the Fourier coefficients of the
shear field with the inside mask $K(\mathbf{n})$ is related to those
without the mask as
\begin{equation}
\tilde{E}_{\mathbf k}^{\rm flat}\pm i\tilde{B}_{\mathbf k}^{\rm flat}
=\sum_\mathbf{k'}
(\hat{E}_\mathbf{k'}^{\rm flat}\pm i\hat{B}_\mathbf{k'}^{\rm flat})~
{}_{\pm 2}W_\mathbf{k'k},
\label{eq:flat_pseudo2}
\end{equation}
where the convolution kernel ${}_{\pm 2}W_{\mathbf{k'k}}$ is
given in terms of the Fourier transform of the mask function,
$K_{\mathbf{k}}$, as
\begin{equation}
{}_{\pm 2}W_\mathbf{k'k}
\equiv L^{-2}K_{\mathbf{k-k'}}
e^{\pm 2i\varphi_\mathbf{k'k}}.
\end{equation}
In the equation~(\ref{eq:flat_pseudo2}) we use the discrete summation,
rather than the infinite-range Fourier transform. The summation runs
over the Fourier-space grids where the resolution of each grid is
specified by the square size $L$ as $(2\pi/L)^2$. The number of grids,
$N_{\rm grid}^2$, needs to be specified by an observer to achieve the
desired coverage of multipole range.

The pseudo spectra $\tilde{\mathbf C}_{\mathbf k}^{\rm flat}$ for the
masked field inside the square boundary is
\begin{equation}
\label{eq:flat_pseudocl2}
\tilde{\mathbf C}_{\mathbf k}^{\rm flat}=
\sum_{\mathbf k'} {\mathbf M}_{\mathbf{kk'}}^{\cal K}
\hat{\mathbf C}_\mathbf{k'}^{\rm flat},
\end{equation}
The mode-coupling matrix for the masking effect is given by
\begin{eqnarray}
M^{\cal{K} \it{EE,EE}}_\mathbf{kk'}&=&
M_\mathbf{kk'}^{\cal{K} \it{BB,BB}}= L^{-2}{\cal
K}_\mathbf{k-k'}\cos^2(2\varphi_\mathbf{k'k}), \\
M^{\cal{K} \it{EE,BB}}_\mathbf{kk'}&=&
M_\mathbf{kk'}^{\cal{K} \it{BB,EE}}= L^{-2}{\cal
K}_\mathbf{k-k'}\sin^2(2\varphi_\mathbf{k'k}), \\
M^{\cal{K} \it{EB,EB}}_\mathbf{kk'}&=& L^{-2}{\cal
K}_\mathbf{k-k'} [\cos^2(2\varphi_\mathbf{k'k})
-\sin^2(2\varphi_\mathbf{k'k})]
\end{eqnarray}
where ${\cal K}_\mathbf{k}$ is the power spectrum of the mask field
inside the square boundary
\begin{equation}
\langle K_\mathbf{k}K^\ast_\mathbf{k'} \rangle 
\equiv L^2\delta^K_\mathbf{k-k'}{\cal K}_\mathbf{k}.
\end{equation}
Combining the equations (\ref{eq:flat_pseudocl1}) and
(\ref{eq:flat_pseudocl2}), the pseudo spectra
$\tilde\mathbf{C}_\mathbf{k}^{\rm flat}$ is related to the true
underlying spectra as
\begin{equation}
\tilde\mathbf{C}_\mathbf{k}^{\rm flat}=
\sum_\mathbf{k''}\mathbf{M}_\mathbf{kk''}^{\cal K}
\sum_\mathbf{k'}\mathbf{M}_\mathbf{k''k'}^{\cal S}
\mathbf{C}_{k'}^{\rm flat}.
\end{equation}

We compute the binned power spectra:
\begin{equation}
\tilde{\cal C}_b^{\rm flat} \equiv \sum_{\mathbf k}^{k\in b}
P_{bk}\tilde{\mathbf{C}}_{\mathbf k}^{\rm flat},
\end{equation}
where the summation corresponds to an azimuthal angle average of
${\mathbf k}$, and runs over all the Fourier modes satisfying the
condition $|\mathbf{k}|\in [k_b,k_{b+1}]$. The operator $P_{bk}$ is
defined to average dimensionless power within the bin of scales
\begin{equation}
P_{bk}=\frac{1}{\nu_b}\frac{k^2}{2\pi},
\end{equation}
and its reciprocal is
\begin{equation}
Q_{kb}=\frac{2\pi}{k^2}.
\end{equation}
The quantity $\nu_b$ is the number of Fourier modes available for the
$b$-th multipole bin, and is approximately given for a mode of $k\gg
1/L$ as
\begin{equation}
\nu_b\equiv\sum_{\mathbf k}^{k\in b}\simeq 2\pi k(k_{\rm
low}^{(b+1)}-k_{\rm low}^{(b)}).
\label{eq:nb}
\end{equation}

Therefore, as in the full-sky approach, the underlying power spectra
can be estimated by deconvolving the pseudo power spectra in the
flat-sky approximation:
\begin{equation}
\mathbf{\cal C}_b^{\rm flat}\simeq (\mathbf{M}^{-1})_{bb'}
\sum_\mathbf{k}^{k\in b'}
P_{b' k}(\tilde{\mathbf C}_{\mathbf k}^{\rm flat}
 -\langle\tilde{\mathbf N}_{\mathbf k}\rangle_{\rm MC}),
\label{eq:flat_deconvCl}
\end{equation}
where the mode mixing matrix is given as
\begin{equation}
\mathbf{M}_\mathbf{bb'}=\sum_{\mathbf{b''}}
\mathbf{M}^{\cal K}_\mathbf{bb''}
\mathbf{M}^{\cal S}_\mathbf{b''b'}. 
\end{equation}
with
\begin{eqnarray}
\mathbf{M}^{\cal S}_{bb'}&=&\sum_{\mathbf k}^{k\in b}P_{bk}
\sum_{\mathbf k'}^{k'\in b'}
\mathbf{M}^{\cal S}_\mathbf{kk'}Q_{k' b'}, \nonumber \\
\mathbf{M}^{\cal K}_{bb'}&=&\sum_{\mathbf k}^{k\in b}P_{bk}
\sum_{\mathbf k'}^{k'\in b'}
\mathbf{M}^{\cal K}_\mathbf{kk'}Q_{k' b'}.
\end{eqnarray}
We use the equation~(\ref{eq:flat_deconvCl}) to reconstruct the power
spectra from the pseudo spectra in the flat-sky approximation.

\section{Simulations of Shear Maps}
\label{sec:simulations}

We use simulated maps of shear fields to test the pseudo-spectrum
technique for reconstructing the shear power spectrum.  We address how
the method works for the case of lensing maps, where the lensing power
spectrum has distinct shapes from the CMB power spectrum: the lensing
spectrum has greater amplitudes at smaller scales.  We also study the
validity of the flat-sky approximation for recovering the input power
spectrum down to very small angular scales in the presence of the
finite-sky survey and masking effects.

\subsection{Simulating the Lensed Sky}
\label{subsec:simulations}

We construct two types of simulated shear fields: flat-sky
non-Gaussian fields from ray-tracing simulations to test the flat-sky
method and full-sky Gaussian fields to test the full-sky deconvolution
method.

\subsubsection{Full Simulations in Flat-Sky Limit}
\label{subsubsec:nbody}

We use the shear maps constructed from ray-tracing simulations
developed by \citet{Sato09} and test the flat-sky deconvolution
method. The ray-tracing simulations are performed by placing N-body
simulation boxes with two different sizes (240$h^{-1}$Mpc and
480$h^{-1}$Mpc at a side length) to cover a light cone of angular size
$5^\circ\times 5^\circ$ from $z$=0 to 3.5. The N-body simulations are
based on the parallel Tree-Particle Mesh code Gadget-2
\citep{Springel05}. The ray-tracing simulation follows the standard
multiple lens plane algorithm \citep{JSW00,Hamana01}: separating the
comoving distance between an observer and a source by fixed intervals
of 120$h^{-1}$Mpc, computing the projected density contrast at each
lens plane along the line-of-sight, we obtain the two-dimensional
gravitational potential field related to the density contrast via
Poisson equation.  In this paper we use the simulations with a single
source redshift $z=1$.  The initial ray directions are set on
2048$\times$ 2048 Cartesian grids (resolution $300/2048\simeq 0.15$
[arcmin]) and the ray positions at each lens plane are traced backward
to the source plane via the lens equation. We obtain the shear field
from the Jacobian matrix describing deformation of an infinitesimal
light ray bundle.  It should be noted that the obtained shear
field does not obey the periodic boundary condition due to the
projection to a single observer point through the different simulation
boxes. The fundamental mode of the shear field is given by $l_{\rm
f}=2\pi/L=72$ with $L=(5/180)\pi$. We use 1000 realizations of the
shear maps in order to sufficiently reduce statistical scatters in the
measured power spectra.

We adopt the concordance $\Lambda$CDM model: the present-day density
parameters of matter, cosmological constant, and baryon are
$\Omega_{\rm m}=0.238$, $\Omega_\Lambda=0.762$, and $\Omega_{\rm
b}=0.042$, respectively; the Hubble parameter $H_0=73.2~{\rm
km~s^{-1}~Mpc^{-1}}$; the spectral index $n_s=0.958$; the rms density
fluctuations in a sphere of 8$h^{-1}$Mpc radius $\sigma_8=0.76$ used
for the power spectrum normalization \citep{Spergeletal07}.

\subsubsection{Gaussian Simulations in Full-Sky Limit}
\label{subsubsec:gauss}

To test the full-sky formalism (Sec. \ref{subsec:full}), we construct
Gaussian shear fields using the HEALPix software \citep{Gorski05}. We
assume the survey geometry to be a rectangular shape covering 2000
square degree (see the top panel of Fig.~\ref{fig:mask}). The
resolution parameter of HEALPix pixelization $N_{\rm side}$ is set to
be 1024 corresponding to $\sim 4$ arcmin for the pixel scale and the
maximum multipole $l_{\rm max}=2000$.

The spectrum of the shear fields is of $E$-mode only assuming the
concordance cosmological model same as the N-body simulations
(Sec.~\ref{subsubsec:nbody}). In the weak lensing limit, the $E$-mode
shear power spectrum $C_l^{EE}$ is equivalent to the power spectrum of
lensing convergence field, the weighted mass density field projected
between source galaxies and an observer along the
line-of-sight. Assuming the Limber's approximation \citep{Limber54},
the $E$-mode power spectrum is simply expressed as a weighted
projection of the the matter power spectrum $P_\delta$(k):
\begin{eqnarray}
C_l^{EE} =\int_0^{\infty}\!dr~ r^{-2} W_{\rm GL}^2(r)
P_\delta\!\left(k=\frac{l}{r}; z\right),
\label{eq:kappa}
\end{eqnarray}
where $r$ is the comoving angular diameter distance, and we throughout
this paper assume a flat universe. Note that the comoving distance is
given as a function of redshift, i.e. the distance-redshift relation
$r=r(z)$.
The lensing efficiency function, 
$W_{\rm GL}(r)$, is given by
\begin{equation}
W_{\rm GL}(r)\equiv
\frac{3}{2}\left(
\frac{H_0}{c}
\right)^2\Omega_{\rm m}a^{-1} r
\int^{\infty}_r\!dr' 
G(r')\frac{r' -r}{r'},
\end{equation}
where $G(r)dr=p(z)dz$ is the normalized source distribution and 
\begin{equation}
p(z)\propto \delta_D(z-z_s)
\end{equation}
with $z_s=1$.  

\subsection{Simulating Realistic Masks}

\begin{figure}
\begin{center}
\hspace{1.2cm}
\includegraphics[width=5.5cm]{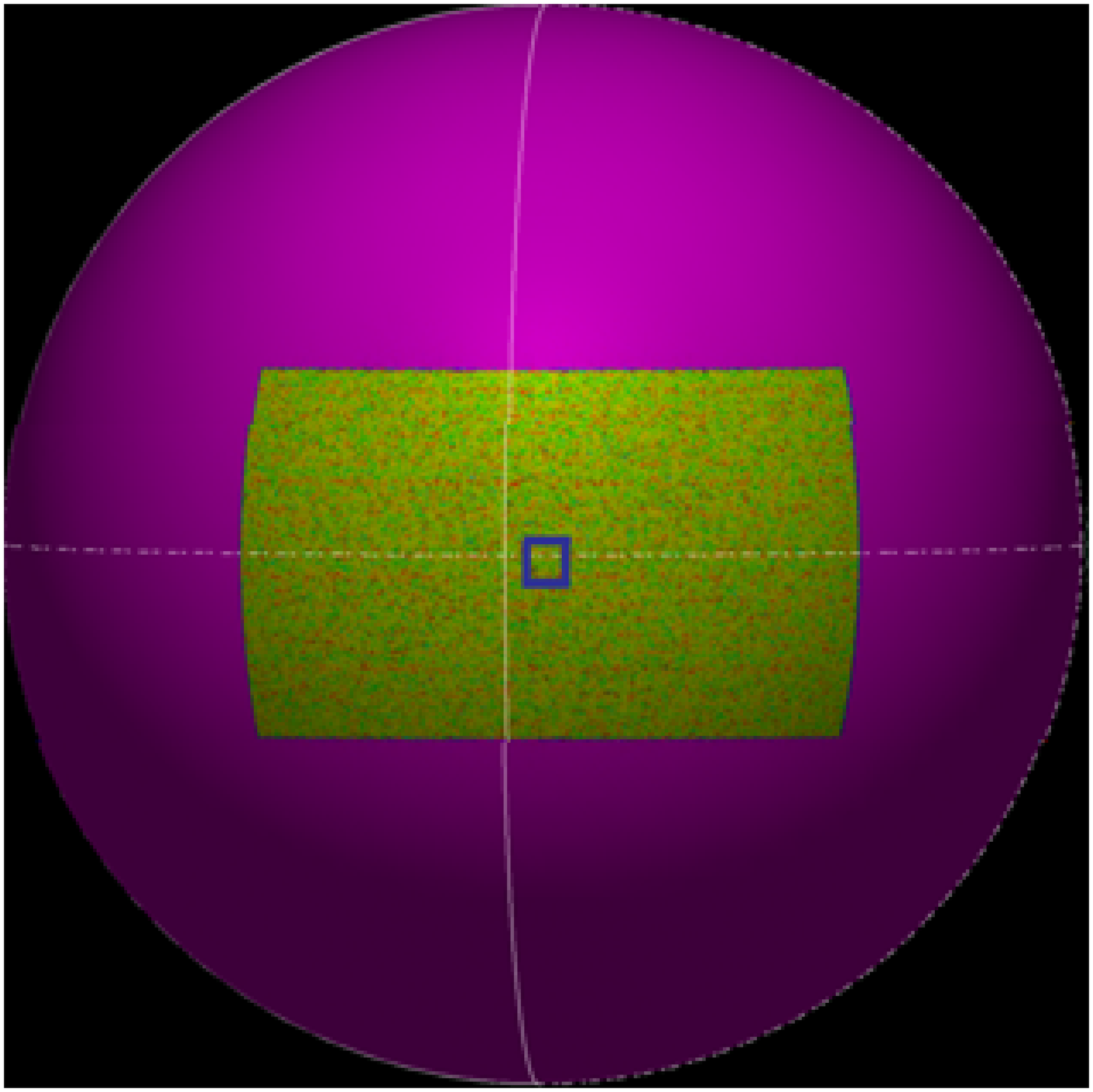}
\includegraphics[width=7.3cm]{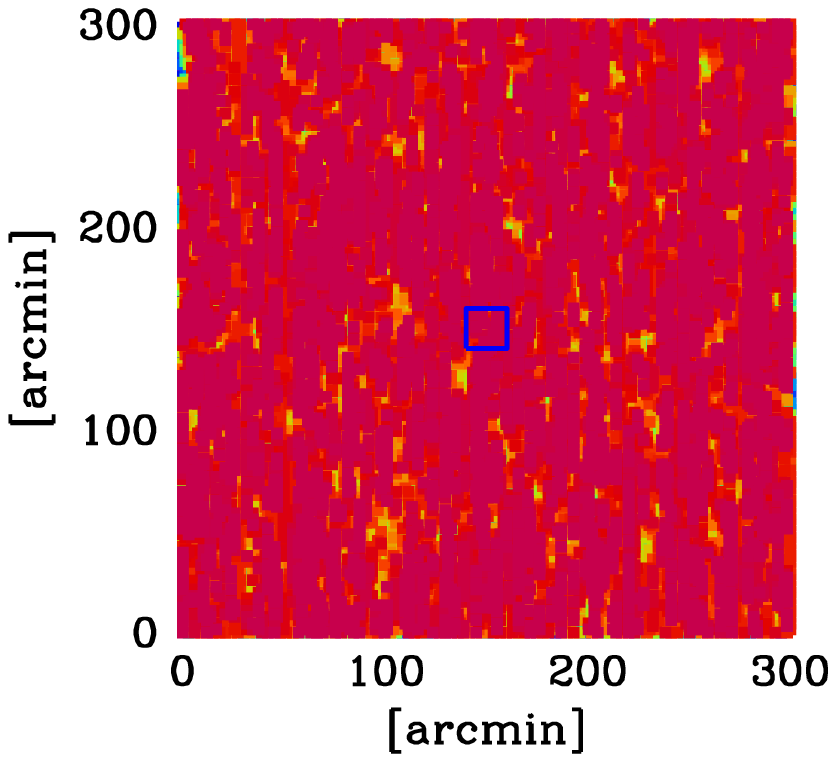}
\vspace{-0.5cm}
\includegraphics[width=6.8cm]{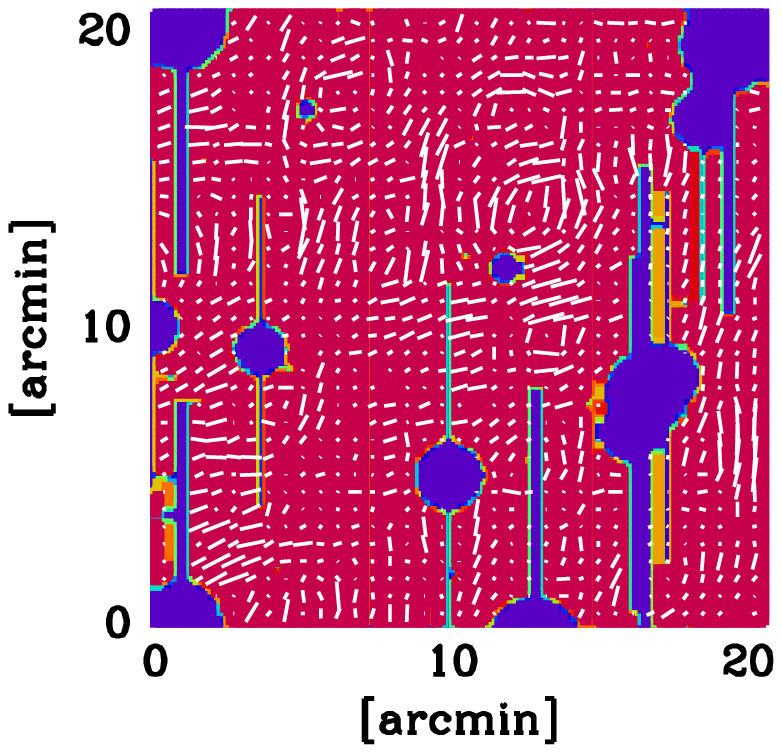}
\vspace{0.5cm}
\caption{{\em Top panel}: A representative realization of the full-sky
 simulations assuming that the shear field is Gaussian. The simulated
 survey geometry is a rectangular shape with 2000 sq. deg. area. {\em
 Middle panel}: Similarly, but for the flat-sky approximation, which
 is done by ray-tracing simulations based on N-body simulations for
 the concordance $\Lambda$CDM model. The simulated map has a square
 shape with 25 sq. deg. area (the corresponding area is displayed by
 the square line in the upper panel). {\em Bottom panel}: A typical
 configuration of masked regions in a patch of $20\times 20$
 arcmin$^2$ (the corresponding area is displayed by the square line in
 the middle panels). The masked regions occupy about 25\% of the total
 simulated area.}
\label{fig:mask}
\end{center}
\end{figure}

We consider three types of masks typical for ground-based imaging data
analyses: point sources, saturation spikes, and bad pixels. First
masked regions have circular shapes to model ``bright stars'' that are
randomly distributed in the shear map. The radii of the bright star
masks are randomly chosen ranging from 0.2 to 2 arcminutes. When the
radius of the circle is greater than 1 arcmin, we add a
rectangular-shape mask to model ``saturation spikes'' around a bright
star. The mask has the size of $0.2r\times 5r$ ($r$ is the radius of
masked circle) with the same center as the circle. The different
masked regions are allowed to be overlapped. We also mask all the
pixels in a $y$-direction row randomly selected with 5\% probability.
About 25\% fraction of the area is totally masked in each simulation
realization, which is typical for a ground-based imaging survey such
as the Subaru Telescope \citep{Hamana03}. The configuration of masked
regions is demonstrated in the bottom panel of Fig.~\ref{fig:mask}.

Since the pixel size of the ray-tracing simulations (0.15 arcmin) is
smaller than the bright star mask, the mask function $K(\mathbf n)$ is
set as $K=0$ when the center of the pixel is masked, otherwise 1.  The
middle panel of Fig.~\ref{fig:mask} shows a typical configuration of
the simulated map with mask, while the spatial resolution is degraded
in the plot.  On the other hand, the Gaussian simulations have a large
pixel size (4 arcmin) compared to the bright star mask, the mask
function $K(\mathbf n)$ has a fractional value. Besides this
difference the masking configuration is kept same in all the
realizations for both the full-sky and flat-sky simulations.

\subsection{Intrinsic Ellipticities}

The intrinsic ellipticities of source galaxy shapes generate a white
noise contamination to the power spectrum measurement assuming random
orientations between different galaxies:
\begin{equation}
N_l=\frac{\sigma_\epsilon^2}{\bar{n}_g},
\label{eq:shot}
\end{equation}
where $\sigma_\epsilon$ is the rms ellipticities per component and
$\bar{n}_g$ is the mean number density of source galaxies. In the
following we assume $\sigma_\epsilon=0.22$ and $\bar{n}_g=30$
arcmin$^{-2}$.

After masking the simulated shear fields, we add a Gaussian random
noise with the variance of $\sigma_\epsilon^2/(\bar{n}_{\rm
g}\Omega_{\rm pix})$ to each pixel, where $\Omega_{\rm pix}$ is the
pixel size.  For Gaussian simulations, we add a effective noise with
the variance of $\sigma_\epsilon^2/(f_{\rm pix}\bar{n}_{\rm
g}\Omega_{\rm pix})$ to pixels inside the survey area, where $f_{\rm
pix}$ denotes the unmasked fraction roughly equal to 0.75.

\section{Results}
\label{sec:results}

\subsection{Power Spectrum Reconstruction}

\begin{figure*}
\begin{center}
\includegraphics[width=8cm]{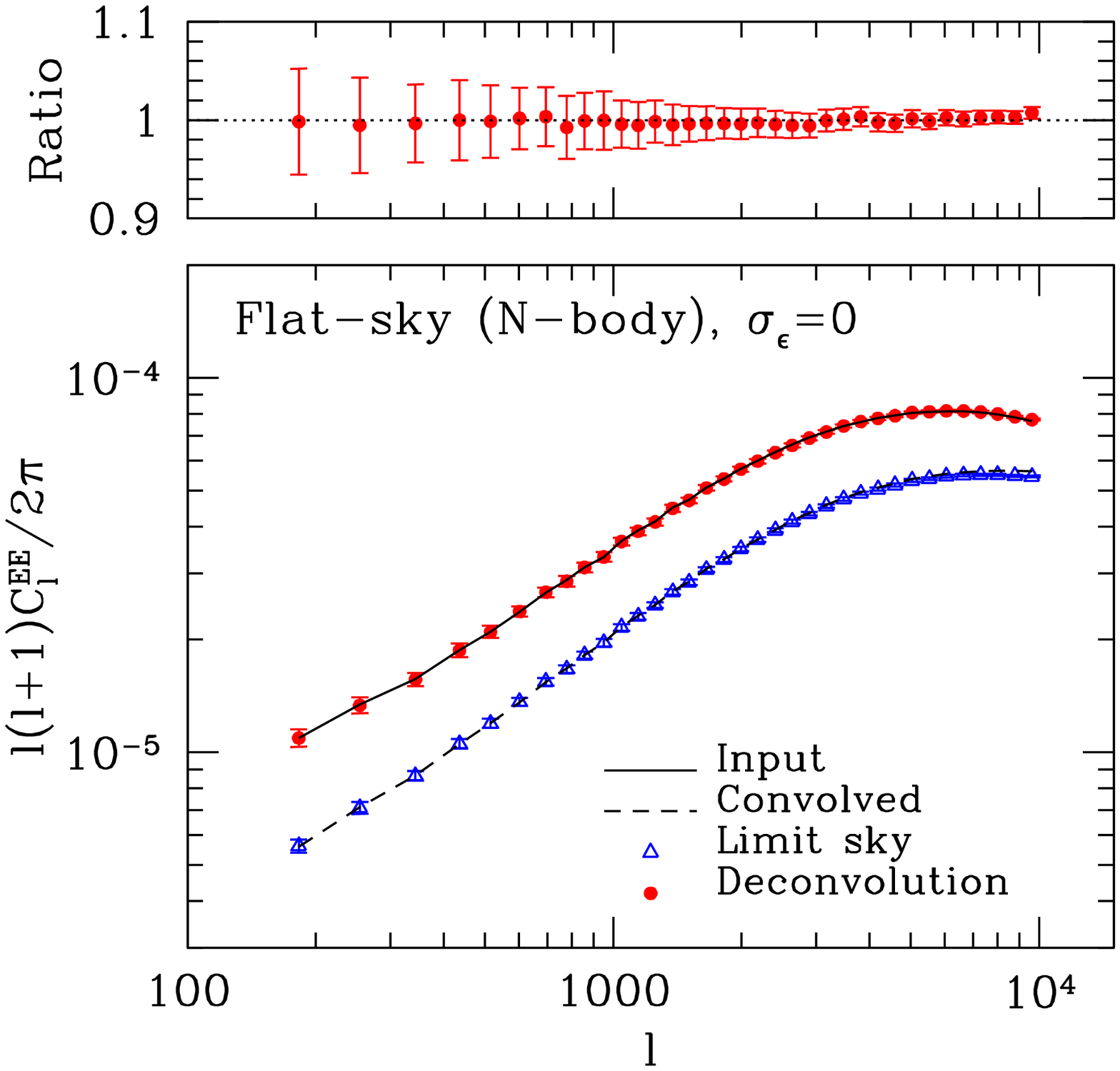}
\includegraphics[width=8cm]{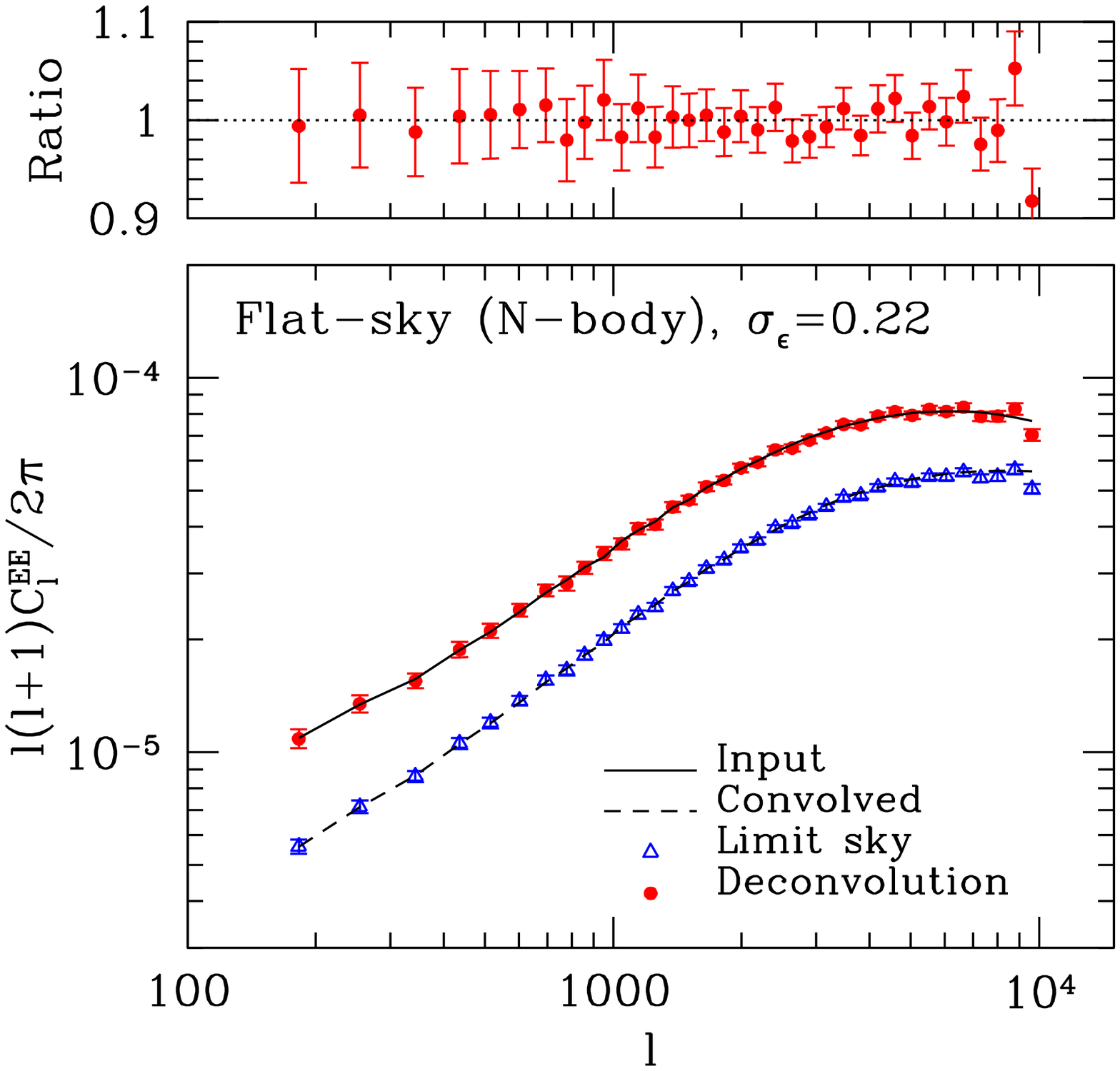}
\includegraphics[width=8cm]{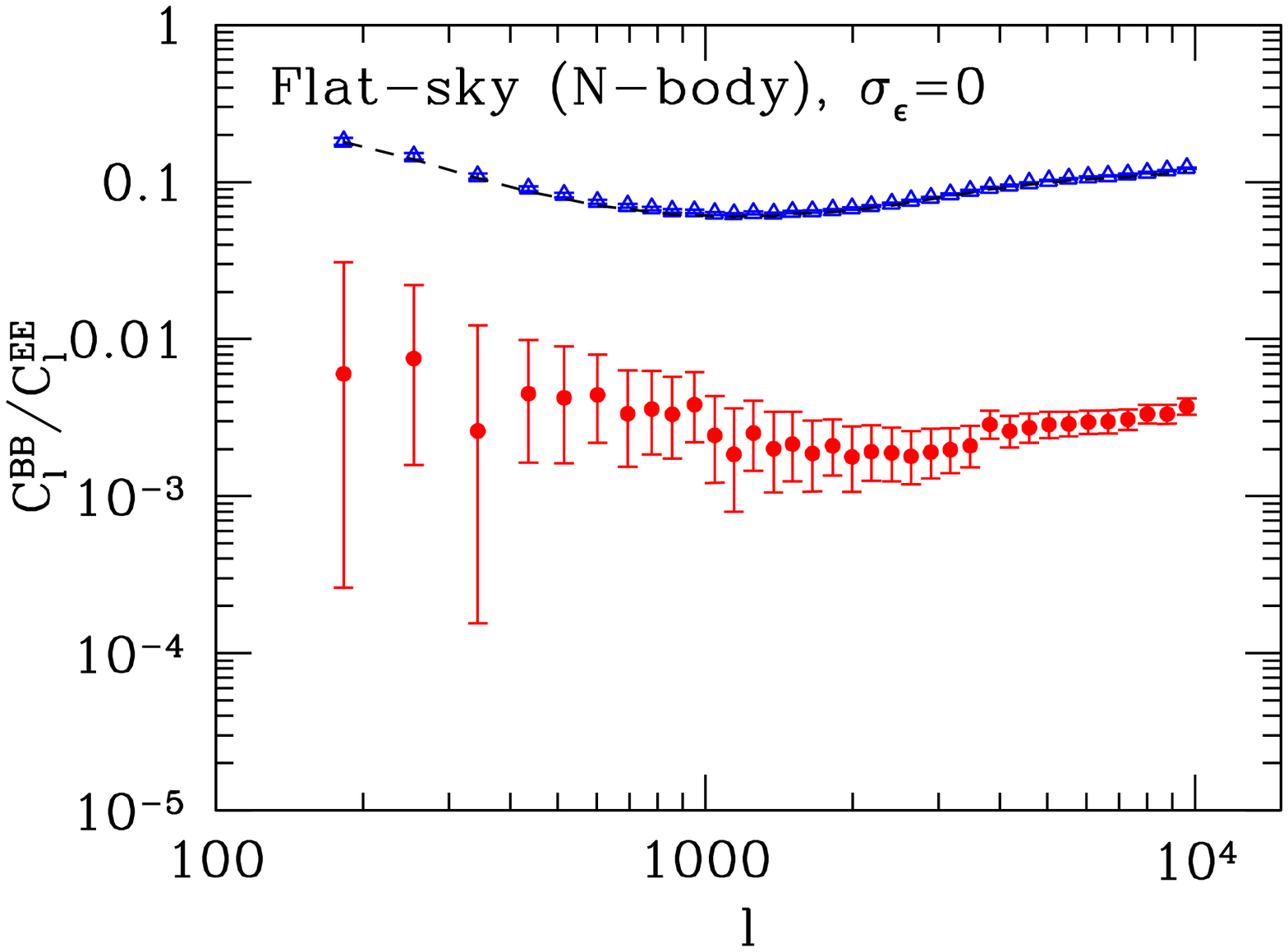}
\includegraphics[width=8cm]{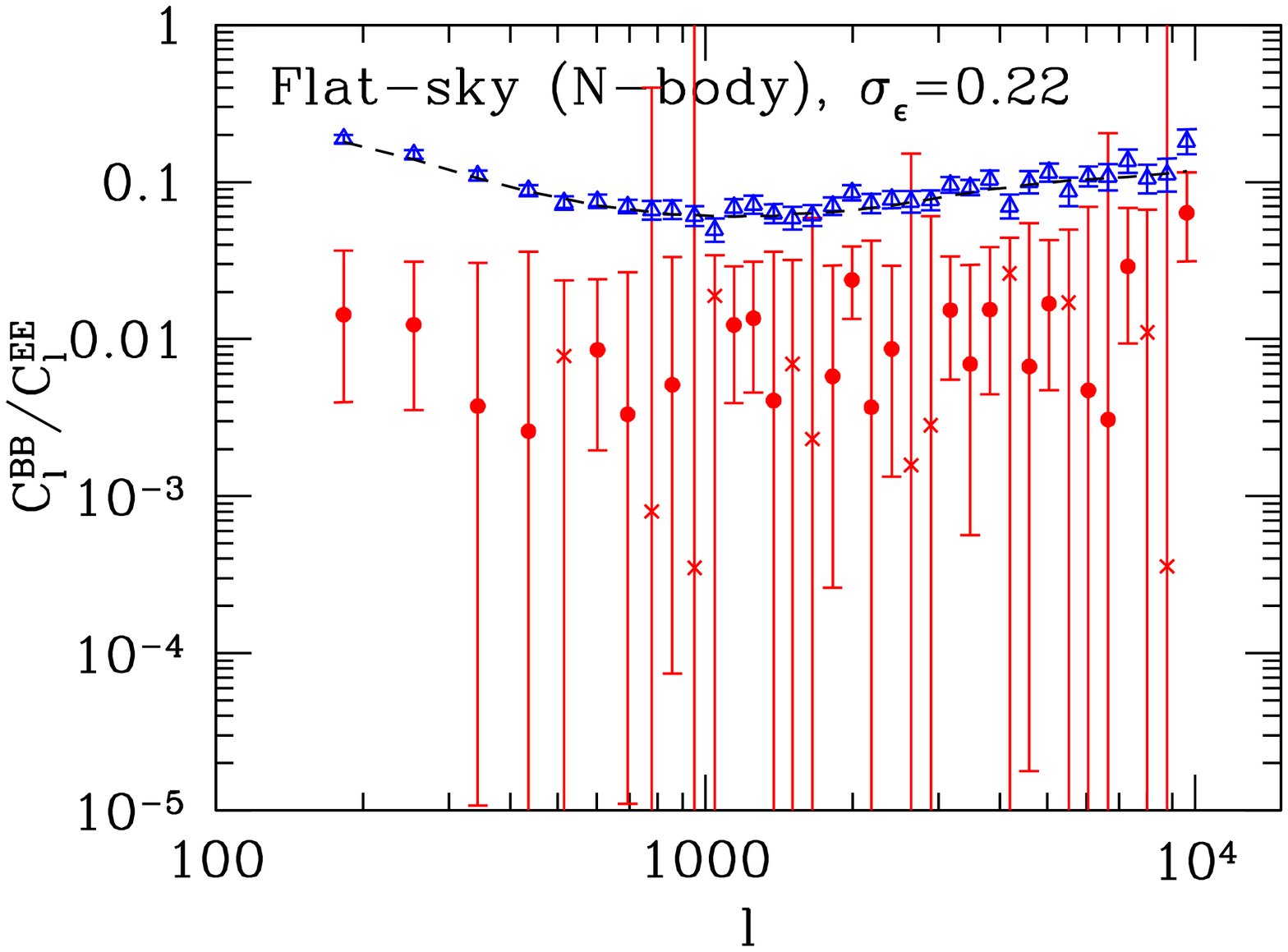}
\caption{Shear power spectra averaged over 1000 realizations of the
ray-tracing simulations. Left panels show the noise-free results,
while right panels show the results with the intrinsic shape noise.
Note that the theoretical prediction for the shot noise power spectrum
(see Eq.~[\ref{eq:shot}]) is subtracted from the measured power
spectra. {\em Middle panel}: The solid lines show the input
convergence power spectrum. The triangle symbols are the pseudo
$E$-mode power spectrum directly measured from the masked simulations
(without correcting for the survey geometry effect), which
significantly underestimate the input power spectrum denoted by the
solid curve. The dashed lines show the theoretical pseudo spectrum
obtained by convolving the input spectrum with the mask. The circle
symbols show the reconstructed $E$-mode power spectra correcting for
the survey geometry effect based on the method in
Sec.~\ref{subsec:flat}. As explicitly shown in the top panel, the
reconstructed $E$-mode spectrum agrees with the input spectrum at a
sub-percent level over a wide range of multipoles down to $\ell =
10^4$. The deviation is much smaller than the expected statistical
errors for a survey of 2000 square degree area coverage denoted by the
error-bars. The errors are estimated by scaling the measured scatters
of the original simulations, which has 25 sq. degree area, with a
factor $\sqrt{2000/25}$.  {\em Bottom panel:} The $B$-mode power
spectra before and after correcting for the survey geometry effects,
compared with the corresponding $E$-mode spectra. Note that the
$B$-mode spectrum can be negative after multiplying the mode-mixing
matrix with the measured $E/B$-mode spectra, and the cross symbols
denote such negative $B$-mode values. The reconstruction method
suppresses the residual $B$-mode below a percent level shown by the
circle symbols.}
\label{fig:nbody}
\end{center}
\end{figure*}

\begin{figure*}
\begin{center}
\includegraphics[width=8cm]{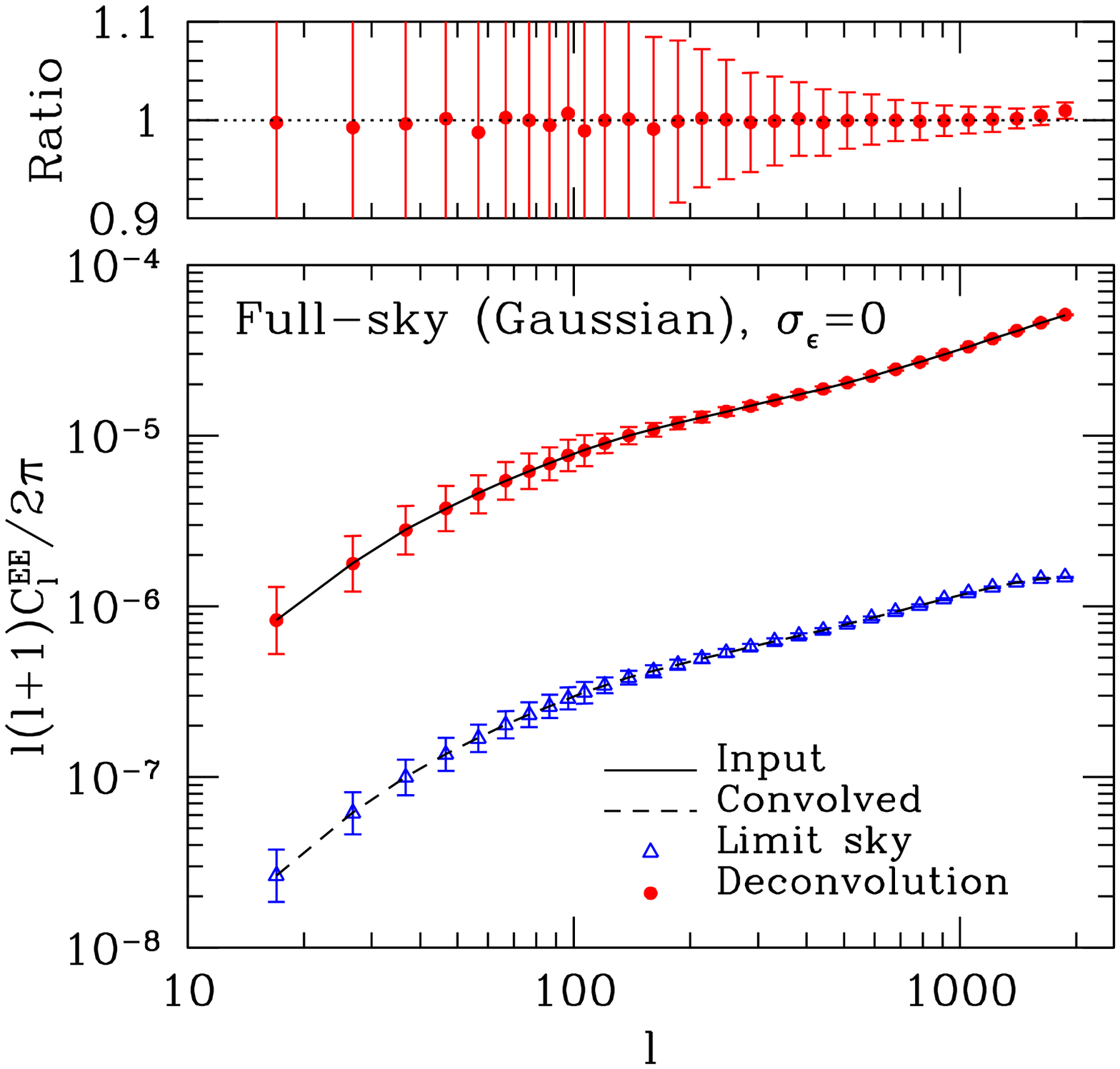}
\includegraphics[width=8cm]{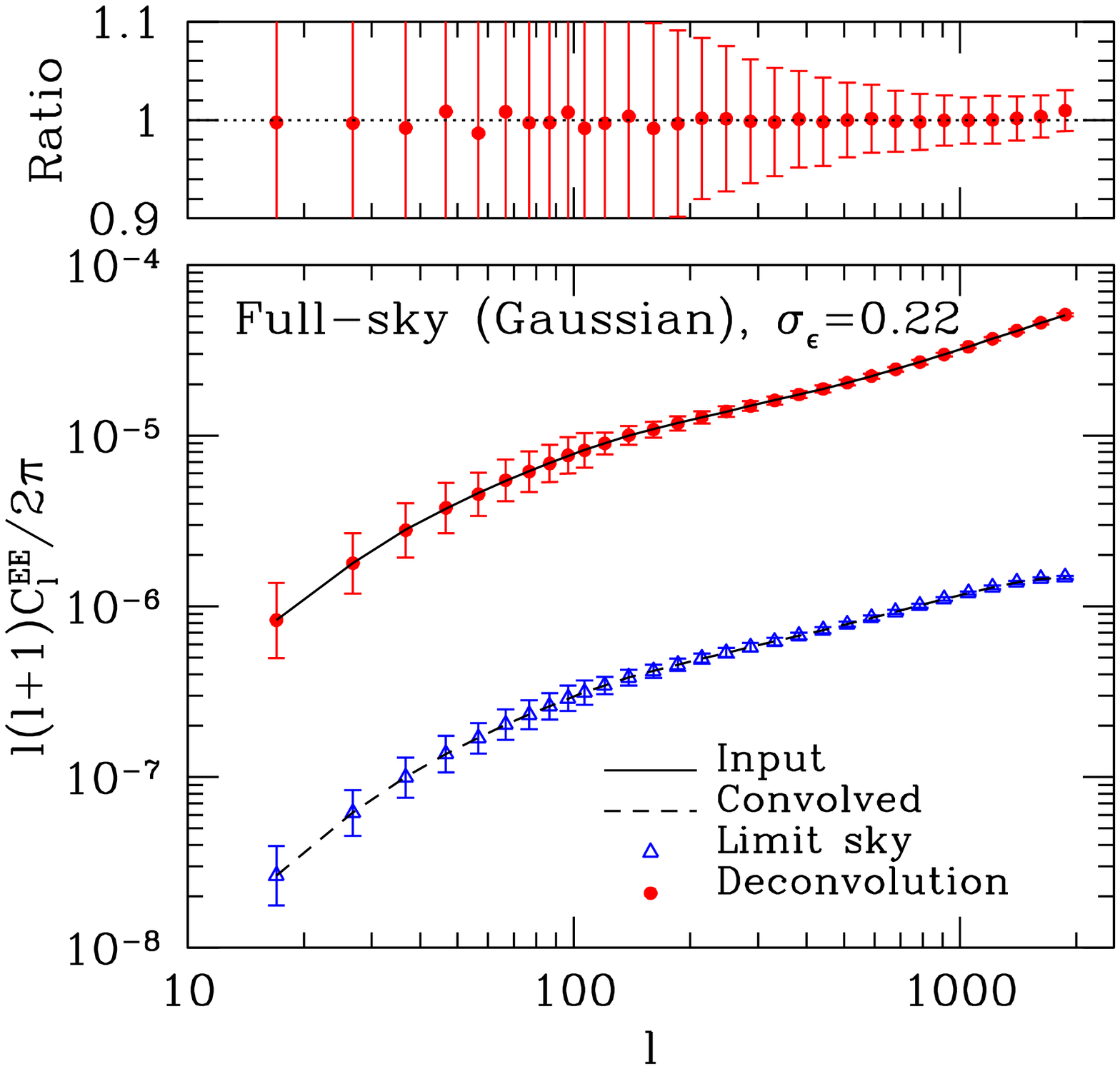}
\includegraphics[width=8cm]{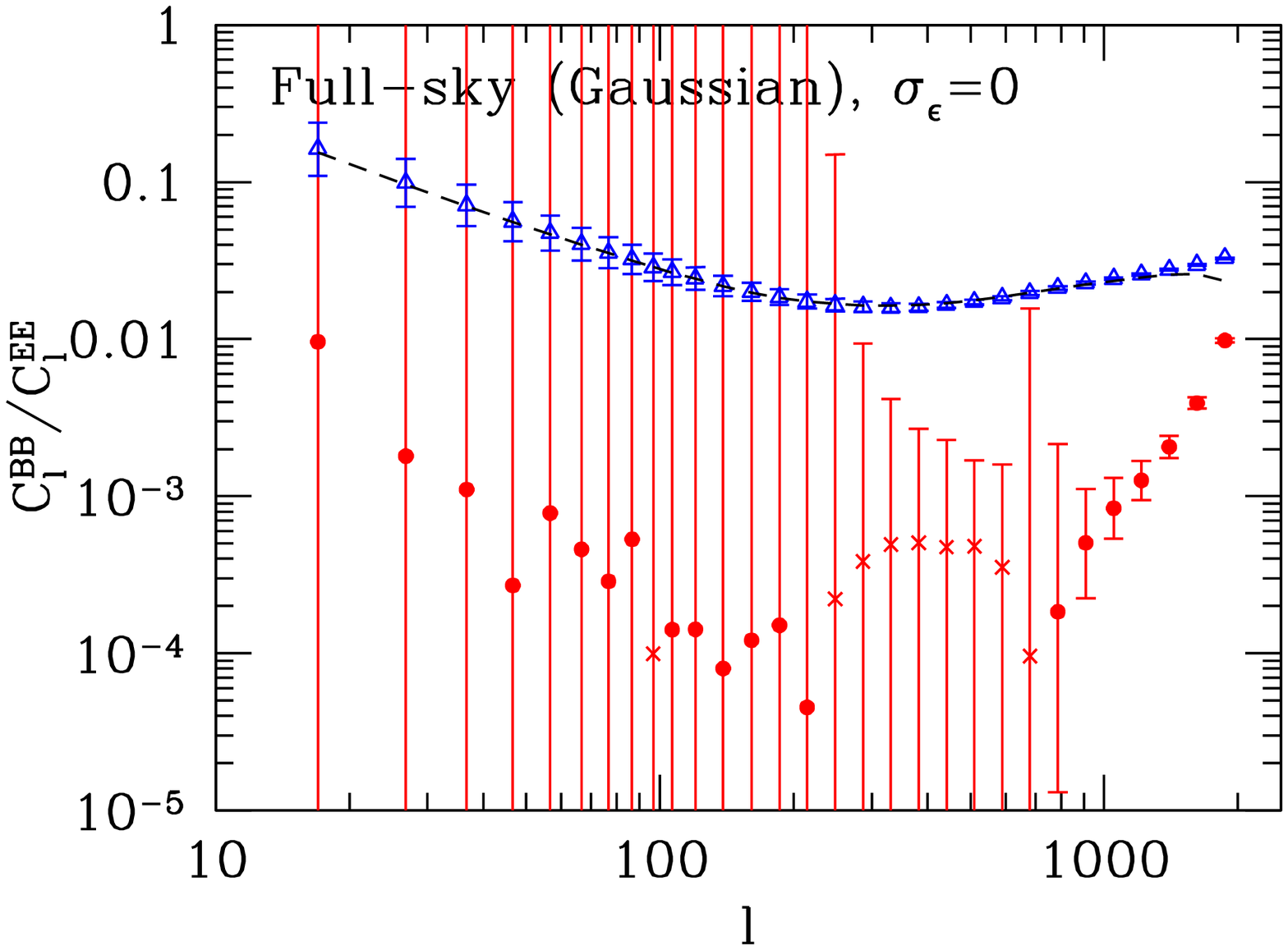}
\includegraphics[width=8cm]{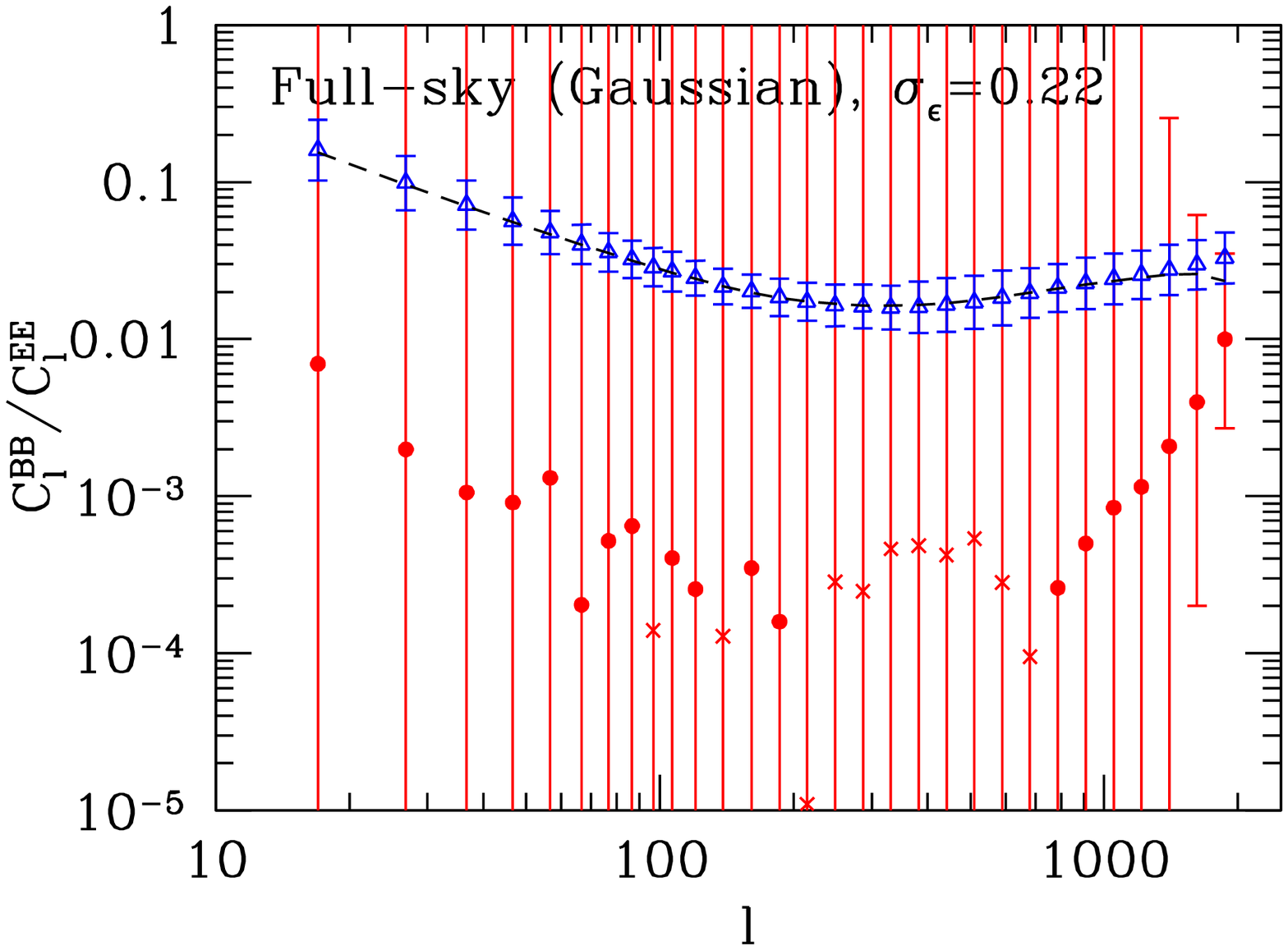}
\caption{Similar to Fig. \ref{fig:nbody} but for the Gaussian
simulations in full-sky approach using 1000 realizations.  Again the
method developed in Sec.~\ref{subsec:full} reconstruct the $E$-mode
power spectrum at a sub-percent level. The systematic error in the
reconstruction is much smaller than the expected statistical errors
for a survey of 2000 sq. deg. coverage, both with (left-panel) and
without (right panel) the intrinsic shape noise.}
\label{fig:gauss}
\end{center}
\end{figure*}

Fig.~\ref{fig:nbody} shows the results of the shear spectrum
reconstruction in the flat-sky approximation using 1000 realizations
of the ray-tracing simulation.  The left panel shows the noiseless
results, while the right panel shows the results with intrinsic
ellipticity. The shot noise spectra are subtracted from the measured
spectra using the theoretical prediction (Eq.~[\ref{eq:shot}])) in
this plot. In reality the shot noise spectra can be estimated by
erasing the coherent cosmic shear signals in each galaxy shape,
e.g. by remeasuring the shear spectra after randomly rotating
orientations of each galaxy.

Middle panels in Fig. \ref{fig:nbody} plot the $E$-mode power
spectrum.  The solid curve denotes the input power spectrum, while the
dashed curve is the convolved spectrum, i.e. the measured power
spectrum from the simulated shear maps without correcting for the
survey geometry effect. The convolved spectrum systematically
underestimates the underlying amplitudes over all the scales.  On
angular scales larger than a typical scale of masked areas (small
$l$), the amplitude offset is roughly determined by the square of the
masked area fraction i.e. $0.75\times0.75\simeq 0.56$. On the other
hand, the offset on larger $l$ is given by the area fraction,
$0.75$. The modes, which angular scale is larger than the mask
size, is fractionally affected by the mask and thus the amplitude
decreases proportional to the masked fraction. Their power is thereby
proportional to the square of the unmasked fraction. On the other
hand, when the scale of modes is mush smaller than the mask size,
their amplitude is 0 in masked region or 1 in unmasked region. As a
result, their square value also has a value of 0 or 1 and thus the
power is proportional just to the unmasked fraction (not the square).

The triangle symbols show the mean pseudo power spectra over 1000
realizations of the shear maps and the circle symbols are the
deconvolved one. The error-bars represent the expected $1\sigma$
statistical errors of the band powers for a 2000 square degree
survey. Since the ray-tracing simulation has the area of 25 square
degree, we estimate the errors by multiplying the measured scatters
among 1000 realizations with a factor $\sqrt{2000/25}$. We set the
binning width wider than the fundamental mode $l_{\rm f}=2\pi/L=72$:
the first 10 bins have linearly equal spacing as $l_{\rm
min}^{(b)}=0.6l_f+1.2l_fb$, while the remaining 25 bins have
logarithmically equal spacing up to $l_{\rm max}=10000$.  Our
reconstruction method successfully recovers the input $E$-mode
spectrum for both cases without intrinsic noise (left panels) and with
noise (right panels). The top panel explicitly shows the ratio of the
reconstructed $E$-mode power spectra to the input spectrum.  The
accuracy of the reconstruction achieves a sub-percent level over the
most ranges of multipoles.

The lower panels show the $B$-mode power spectrum relative to the
$E$-mode spectrum. The mask and finite-sky effect generates $B$-mode
due to the mode mixing. In fact, there is a significant leakage of the
$E$-mode into the $B$-mode ($\sim$10\% of E-mode power) as shown in
triangle symbols. We find that the leakage disappears to be less than
one percent after deconvolution (see the filled circle symbols).

Fig.~\ref{fig:gauss} shows the results of the full-sky approach
(Sec.~\ref{subsec:full}) using 1000 Monte Carlo simulations of
Gaussian shear fields. The error-bars represent $1\sigma$ statistical
dispersion for a survey with 2000 square degree coverage.  The binning
width is set to be wider than $l_{\rm f}=\sqrt{\pi/f_{\rm sky}}\sim
8$: the first 10 bins have linearly equal spacing as $l_{\rm
min}^{(b)}=2+10\times b$ with, while the remaining 20 bins have
logarithmically equal spacing up to $l_{\rm max}=2000$. The power at
the lowest bin ($12\le l<22$) is affected by the unseen modes larger
than the surveyed region and thus the reconstructed power at the
lowest bin varies with the chosen minimum $l$ and the binning width.
When the minimum $l$ is taken to be slightly larger than $l_{\rm f}$,
we find that the reconstructed power becomes almost equal to the input
power. The accuracy is limited by the statistical errors due to
insufficient sampling of such large-angle modes.  The power spectrum
without correcting for a finite-sky effect (dashed line) is smaller
than the input power spectrum (solid line). The amplitude offset is
roughly determined by the square of unmasked fraction times the survey
area fraction, $0.75\times 0.75\times 2000$ square degree divided by
$4\pi$ steradian, that is about $0.028$.

Summarizing the results in Figs.~\ref{fig:nbody} and \ref{fig:gauss}
the pseudo-spectrum method recovers the input power spectrum at a
sub-percent accuracy over a wide range of multipoles down to $l\simeq
10^4$. As seen from the top panels, the ratio between the true and
reconstructed spectra is close to unity over a range of multipoles we
study. The scatters of the ratio from unity imply possible biases in
the $E$-mode power spectrum reconstruction at each multipole bin, and
are smaller than the statistical errors of power spectrum measurement
expected for a survey of 2000 sq. deg. coverage.  The residual
$B$-mode power spectrum is suppressed below a percent level.

\subsection{Correlated Spectrum Errors due to the Mode Mixing}

Finite-sky effect makes the errors of the shear power spectra
correlate between different bins. In this subsection we study this
effect using the simulated maps.

Let's start our discussion using Gaussian fields.  The statistical
errors of the binned power spectrum is given, in the absence of shot
noise, as
\begin{equation}
\Delta C_b^{({\rm Gauss})}=C_b\sqrt{\frac{2}{f_{\rm sky}\nu_b}},
\label{eq:error}
\end{equation}
where $f_{\rm sky}$ is the sky fraction of survey area and $\nu_b$ is
the number of modes available around the $b$-th bin.

\begin{figure*}
\begin{center}
\includegraphics[width=8cm]{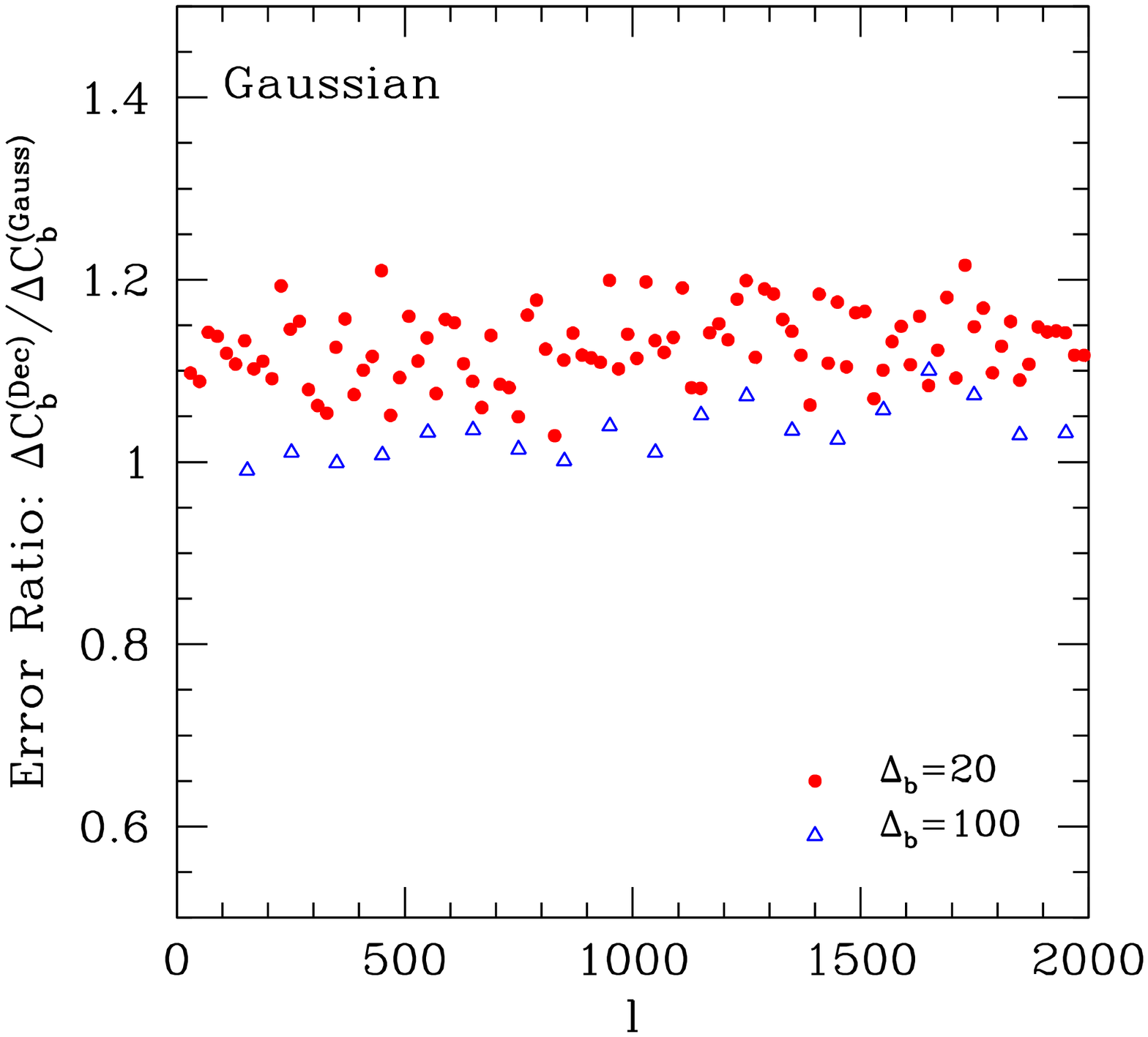}
\includegraphics[width=8cm]{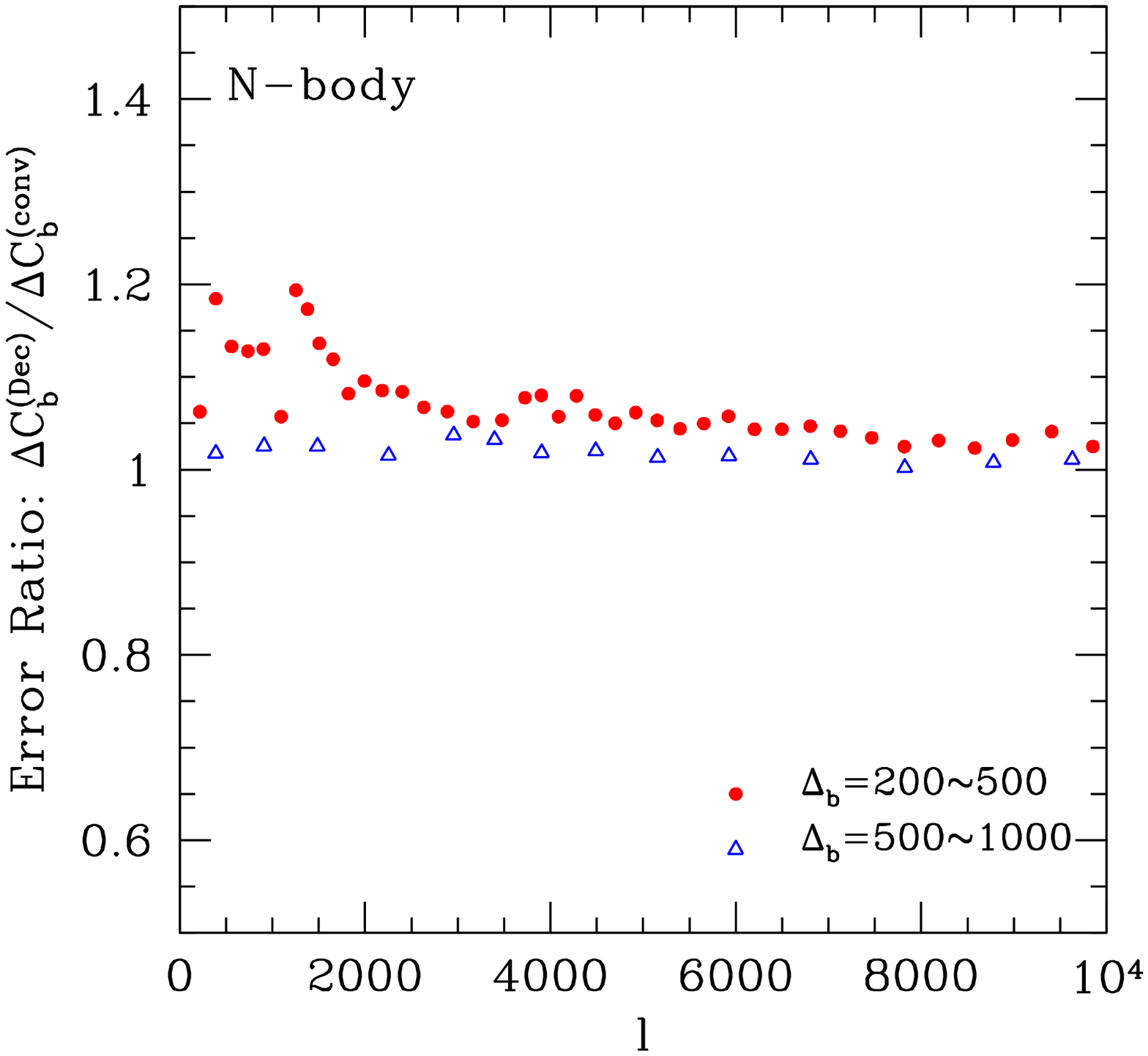} 
\caption{{\it Left panel:} Error increment of the reconstructed shear
power spectrum due to the effects of the finite sky area and the
mask. The errors are measured from 1000 realizations of the full-sky
Gaussian simulations and then compared with the Gaussian error
expectation (Eq.[\ref{eq:error}]). The survey geometry effect
increases the power spectrum errors, but it becomes insignificant at a
wider bin width $\Delta l=100$, about factor 10 wider than the
fundamental mode of the 2000 sq. deg. survey area, $l_{\rm f}\simeq
8$. {\em Right panel}: Similar to the left panel but for the full
simulations in a flat approximation. The error of the reconstructed
power is compared with that of the unmasked convergence power of the
ray-tracing simulations including the non-Gaussian errors. Again the
mask effect increases the errors, which is suppressed at a wider bin
width. Note that the non-Gaussian errors do not depend on the bin
width.}
\label{fig:error}
\end{center}
\end{figure*}

The left panel of Fig.~\ref{fig:error} compares the Gaussian
expectation (\ref{eq:error}) with the statistical errors in the
$E$-mode power spectrum of 1000 Gaussian shear fields. When the
multipole bin width is taken as $\Delta l=20$, 2.5 times wider than
the the fundamental mode $l_{\rm f}\equiv \sqrt{\pi/f_{\rm sky}}\simeq
8$, the errors for the simulation results are greater than the
Gaussian expectation. The limit of sky area and mask causes correlated
errors between neighboring multipole bins and increases the errors by
$10$-$20$\% over a range of multipoles. When the bin width is enough
wide such as $\Delta l=100\simgt 12.5l_{\rm f}$, the power spectrum
errors become similar to the Gaussian expectations, and the spectra at
different bins become nearly independent.

The right panel of Fig.~\ref{fig:error} compares the error of the
reconstructed shear power relative to the unmasked convergence power
using the ray-tracing simulations, which contain the non-Gaussian
error due to nonlinear clustering in structure formation.  At a bin
width $\Delta l=200$, a factor 3 wider than the fundamental mode
$l_{\rm f}=72$, the power spectrum errors increase by about 5\% on
small multipoles, while the correlated errors are well suppressed at a
wider binning $\Delta l=1000\sim 10l_{\rm f}$. Note that the Gaussian
errors depend on the bin width of multipoles, while the non-Gaussian
errors do not: a wider bin relatively suppresses the Gaussian error
contribution.  As studied in \citet{TakadaJain09} \citep[also
see][]{Sato09}, the non-Gaussian error contribution is significant at
multipoles $\ell \simgt 1000$ for the $\Lambda$CDM model.

Since most of useful cosmological information in the shear power
spectrum resides in the modes around $\ell\simeq 1000$ in the presence
of the shot noise, \cite{TakadaJain09} showed that even a factor 2
increase in the power spectrum errors at the multipoles degrades
accuracies of cosmological parameters only by 10-20\%. Therefore, the
correlated errors due to the survey geometry effect have an
insignificant impact on parameter estimation.

\section{Summary and Discussion}
\label{sec:summary}

We develop a pseudo-spectrum method for reconstructing the cosmic
shear power spectra from actual lensing data. The observed shear field
is limited to be a finite patch of sky and furthermore roughly 25\% of
the survey area is masked due to bright stars.  We apply for the first
time the pseudo-spectrum technique developed in CMB studies to the
lensing field and show that our method successfully recovers the shear
spectra over a wide range of multipoles from $100$ to $10^4$ in both
full- and flat-sky approaches.

We test the flat-sky method using ray-tracing simulations assuming a
square-shaped survey region.  The 25\% fraction of the total area is
masked by realistic configurations for a ground-based survey: circular
shapes for bright stars; rectangular shapes for bright star spikes;
zero padding in one direction for bad pixels. We show that both the
full- and flat-sky methods reconstruct the input $E$-mode power
spectrum in sub-percent accuracy and the residuals are much smaller
than the statistical errors of a power spectrum measurement for a
survey of 2000 square degree coverage. The residual $B$-mode power
spectrum from the $E/B$-mode mixing due to the imperfect correction of
survey geometry is also suppressed below a percent of the $E$-mode
power spectrum.  Our method offers a new means of measuring the cosmic
shear correlations and separating the $E/B$ modes from an actual
survey data.

Although the pseudo-spectrum technique is promising, our method still
yields sub-percent residual of $B$-mode in the reconstructed power
spectra. To further suppress the $B$-mode spectrum, one has to
remove ``ambiguous mode'' that is inevitably generated in a finite
patch of sky \citep{Bunn03}. The ambiguous mode satisfies both the
$E$-mode (rotation-free) and the $B$-mode conditions (divergence-free)
and thus contaminates E/B-mode spectrum reconstructed using the simple
pseudo-spectrum method that we adopt. To eliminate such contamination,
\citet{Bunn03} introduces pure E/B modes that is orthogonal to the
ambiguous modes. Pure pseudo $C_l$ estimator and its optimization
technique of sky apodization have been developed
\citep{Smith06,SmithZal07,Grain09,Kim10}. This technique can be
straightforwardly applied to the lensing case.

In this paper we assume that masked regions are uncorrelated with the
cosmological shear field.  In reality the regions with large shear is
preferentially masked: in a region of massive clusters, we cannot
obtain a fair sample of background galaxies in the central region due
to the dense concentration of member galaxies, where the shearing
effect on background galaxies are greater. Masking such a crowded
region may bias the power spectrum measurement.  This contaminating
effect can be estimated by combining ray-tracing simulations with halo
catalogs in the underlying N-body simulations. This is our future
project, and will be presented elsewhere.

\bigskip

We deeply appreciate Masanori~Sato for kindly providing ray-tracing
simulation data. We also thank an anonymous referee for careful
reading and providing useful comments.  C.H. acknowledges support from
a Japan Society for Promotion of Science (JSPS) fellowship.  This work
is in part supported in part by JSPS Core-to-Core Program
``International Research Network for Dark Energy'', by Grant-in-Aid
for Scientific Research from the JSPS Promotion of Science
(18072001,21740202), by Grant-in-Aid for Scientific Research on
Priority Areas No. 467 ``Probing the Dark Energy through an Extremely
Wide \& Deep Survey with Subaru Telescope'', and by World Premier
International Research Center Initiative (WPI Initiative), MEXT,
Japan.


{}

\end{document}